\documentclass[aps,pra,preprint]{revtex4-1}
\usepackage{graphicx}
\usepackage{bm}
\usepackage{mathrsfs}

\begin{document}
\date{\today}

\title{Simple model of photo acoustic system for greenhouse effect}
\author{Akiko Fukuhara}
\email{fukuhara@hit.ac.jp}
\affiliation{Hokkaido Institute of Technology, 7-15 Maeda, Teine Sapporo 006-8585 Japan.}
\author{Fumitoshi Kaneko}
\email{toshi@chem.sci.osaka-u.ac.jp}
\affiliation{Department of Macromolecular Science, Graduate School of Science, 
Osaka university, Toyonaka Osaka 560-0043 Japan.}
\author{Naohisa Ogawa}
\email{ogawanao@hit.ac.jp}
\affiliation{Hokkaido Institute of Technology, 7-15 Maeda, Teine Sapporo 006-8585 Japan.}

\begin{abstract}
The green house effect is caused by the gases which absorb infrared ray (IR) emitted by the earth.
It is worthwhile if we can adjudicate on which gas causes the greenhouse effect in our class.
For this purpose, one of our authors, Kaneko has designed an educational tool 
for testing greenhouse effect \cite{Kaneko}. 
This system (hereafter abbreviated PAS) is constructed based on photo acoustic effect.
Without difficulty and high cost, we can build PAS and check the IR absorption of gas.
In this paper we give the simple theoretical basis for this PAS.
The amplitude of sound observed in PAS depends on the modulation frequency of IR pulse. 
Its dependence can be explained by this simple model.  
Further we show the sound amplitude does not depend  on the thermal diffusion, which provides the accuracy
of amplitude as the IR absorption rate of the gas.
According to this model, sound signal is not the sinusoidal function and it has higher harmonics.
The theory and experiment are compared in the third  harmonics by spectrum analysis.
From this apparatus and theory, students can study not only the greenhouse effect but also the
basics of physics.
\begin{description}
\item[PACS numbers]
01.40.-d, 92.70.-j, *92.30.Np, *43.20.Hq, 33.20.-t, 89.60.Gg
\end{description}
\end{abstract}
\maketitle

\section{Photo acoustic effect}

The history of photo acoustic (PA) effect is very old. 
It was found 130 years ago by G. Bell and discussed by several famous authors \cite{history}.
Recently the Photo Acoustic system (PAS) is renewed by Kaneko et al. as an educational tool 
to show the infrared absorption of greenhouse gases qualitatively \cite{Kaneko}, \cite{GH}.
PAS is shown in figure 1.
The IR source (1) is the heat source, 3 $\Omega$ heating element given 5 V voltage. 
The IR coming out from source passes through the rotating circle with 5 holes (Chopper (2) ), 
and to be a IR pulse. Then the IR pulse  enters into the gas cell (3).
This IR pulse gives  the molecules  vibrational energy,
 and its energy is transformed into the heat of gas when no re-radiation occurs.
 The obtained heat pulses give rise to the pressure oscillation and that gives the sound.
 The sound is transformed into electric signal by microphone (5) 
and its form can be observed on oscilloscope (6).
Two IR windows attached on gas cell are made of $\mbox{KBr}$ with thickness $2 \mbox{mm}$, 
that are not only transparent for IR, but also they work to cut the sound from outside.
As we can see in figure 2, the gas cell (3) has cylindrical form with interior radius $b_0 = 10\mbox{mm}$, length $l_0 = 100\mbox{mm}$, 
and IR passing region has radius $b < 10\mbox{mm}$. (We change the radius $b$ in experiment)
 The CO$_2$ and H$_2$O (Vapor) provide intense sound signals 
but such PA signals are not caused by the major atmospheric components, N$_2$ and O$_2$.
This system clearly demonstrates the origin of the greenhouse effect.

To check the IR absorption of gas directly, it seems plausible to examine the simple measurement of 
temperature of gases under the static IR irradiation, however,
the temperature increase is very small and we cannot obtain the clear result.
The reason is the following.
First, since the pace of temperature increase is too slow, we cannot neglect the effect of thermal diffusion to outside.
Therefore the gas temperature we observe is not determined only by IR absorption of gas.
Second, the effect that the gas container is warmed up by IR and the gas is warmed up secondary by container might be large.  This is related to the material of gas container. 
It is necessary for the gas container that the transmission factor for IR  to be  high enough.
Third, the temperature increase is too small and so the usual thermometer can not  be used.
We need to observe the temperature by using a different physical variable.

The PAS solves these three problems and shows the clear IR absorption for greenhouse gases.
The precise explanation will be given in further section, but we give here some comments.
First, by giving short periodic IR pulse to the gas, we can make the effect of thermal diffusion negligible.
Second, we use $\mbox{KBr}$ which has very high transparency to IR, as gas cell window (4). 
Third, we observe the temperature variation by the pressure variation, that is, the sound. 
Then we have very sensitive observation without time delay. 
From these reasons we can observe the IR absorption of gas very clearly by PAS.


\begin{figure}
\centerline{\includegraphics[width=5cm]{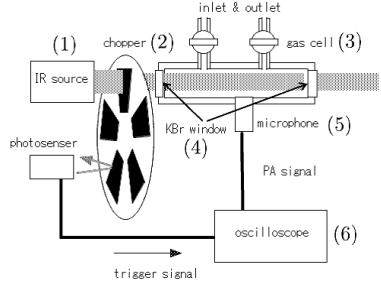}}
\caption{Schematic diagram of a PA experimental system for IR absorption of gases. }
\end{figure}

This PAS provides opportunities to learn the difference in infrared absorption 
between greenhouse gases and other atmospheric gases for high school level student.
However, this educational material is still not enough of a challenge to study quantitative analysis.  
So we need some physical quantities which can be observed easily and that can be compared to simple theory.
For this purpose, we pay an attention to the form and amplitude of sound signal of PA.
From the experiment, we find two facts for the PA signal.
First, though IR pulses are given to the cell, the obtained sound signals are similar to sinusoidal waves.
Second, the amplitude of sound decreases with the increasing modulation frequency of IR pulse.
In this paper we will explain these two phenomena by using simple phenomenological model, 
which makes quantitative analyses possible in this PAS. 
Furthermore we show the sound amplitude does not depend on the thermal diffusion, which provides the accuracy
of amplitude to measure the IR absorption rate of the gas.
In addition we will present an analogy of this PAS to the electric circuit,
which helps students for understanding the PA phenomenon observed in this educational apparatus.

\section{Model}

During the incidental IR beam passes through the gas cell, some of its energy is absorbed into the gas molecules as excitation energy of molecular vibrations. 
Then the molecular internal energy is transported into the kinetic energy (center of mass ), and  increases the gas temperature. 
We define the effective incidental IR power density  $S(\vec{x},t)$[W/m$^3$]  in gas cell, as is explained in Appendix A. 
We further define the temperature field $\theta(\vec{x},t)$ and pressure field $P(\vec{x},t)$ in the gas cell.  Then we obtain two equations for these fields, \cite{theory}

\begin{equation}
\kappa \nabla^2  \theta - \rho_0 ~ c_p ~\frac{\partial\theta}{\partial t}  + S = - \frac{\partial P}{\partial t}, 
\end{equation}

\begin{equation}
\nabla^2 P - \frac{\rho_0}{P_0} \frac{\partial^2 P}{\partial t^2} = 
-\frac{\rho_0}{\theta_0} ~ \frac{\partial^2 \theta}{\partial t^2},
\end{equation}
where we have used the state equation for ideal gas, and the quantities with suffix ``0" means ambient constant quantities. $\kappa$, $\rho$, and  $c_p$ are the heat conductivity of the gas,  mass density, and the specific heat at constant pressure per unit mass  respectively.


\begin{figure}
\centerline{\includegraphics[width=5cm]{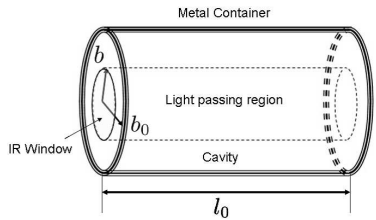}}
\caption{Gas cell has a form of cylinder with inner radius $b_0$ and length $l_0$. 
The circular IR window has radius $b$ with $b \le b_0$.}
\end{figure}

These are simple linear equations, however, it is bit far from intuitive understanding.
So we hope to obtain the simpler model for this PAS which predicts quantitative experimental results.
To carry out this idea, we integrate two equations spatially and we adopt the spatial mean temperature $\bar{\theta}$ rather than the local one $\theta(\vec{x},t)$ as is shown in Appendix B. 
Then we obtain the heat relation.

\begin{equation}
\mathscr{C} \frac{d\bar{\theta}}{dt} = L -  \frac{(\bar{\theta}-\theta_{env})}{\mathscr{R}}, \label{eq:base}
\end{equation}
where the mean temperature of gas $\bar{\theta}$, temperature of environment surrounding cavity
$\theta_{env}$ ($\bar{\theta} > \theta_{env}$), and 
the heat capacity of gas $\mathscr{C}$ (isovolumetric heat capacity).
$\mathscr{R} =  \{2-(b/b_0)^2 \}/(8 \pi l_0 \kappa)$ is the thermal resistance which is explained in Appendix B, where $l_0$ is the length of gas cell, $b_0$ is the inner radius of the gas cell, $b$ is the radius of cell window (See figure 2).
 $L(t)$ is the spacial integration of IR power density $S$, that is, the net energy absorption per second as explained in appendix A.
 
This equation is nothing but the first law of thermodynamics, which means that 
the IR energy given from outside is distributed into the gas and the generated heat diffuses out of cell.
Then the gas temperature increases by the difference of these incoming and outgoing energy.

The function $L(t)$ has a form which is similar to the trapezoidal pulse wave as shown in figure 3 (b). 
The reason is given in the following. At $t=0$, the edge of front window of the cell meets IR coming through a hole 
 region of chopper (figure 4 (a)).
After this moment, a hole region of chopper partially wrap over the window and $L(t)$ becomes larger (figure 4(b)).
At $t=\mu$, the window of the cell is just completely covered by a hole region of chopper 
and $L(t)$ becomes maximum (figure 4(c)).
In a while  $\mu<t<T/2$, $L(t)$ takes maximum value (figure 4 (c)-(e)). From $t=T/2$ to $t=T/2 + \mu$, window of the cell is partially covered by blind region of chopper and $L(t)$ goes down (figure 4 (e)-(g)). 
During the time $T/2 +\mu<t<T$, 
window is covered by blind region of chopper and IR is not coming in (figure 4 (g)-(i)). At $t=T$ the situation is the same as Fig.4 (a).
In this way, the time period $T$  of IR  pulse is corresponding to the angle $\phi_1$, and the time $\mu$ is corresponding to the angle $\phi_2$, where $\phi_1$ is the central angle a set of blind and hole regions of the chopper span, and $\phi_2$ is the central angle the light entrance window of the cell spans (See figure 3(a)). 
So we obtain the relation.

\begin{figure}
\centerline{\includegraphics[width=10cm]{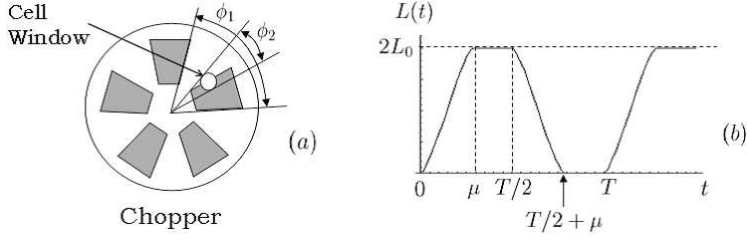}}
\caption{(a) Chopper and front window of the cell, \\
~~(b) IR luminosity into the cell}
\end{figure}

\begin{equation}
\mu = \frac{\phi_2}{\phi_1} ~ T.
\end{equation}


\begin{figure}[htbp]
\begin{center}
\includegraphics[width=0.8\hsize,keepaspectratio]{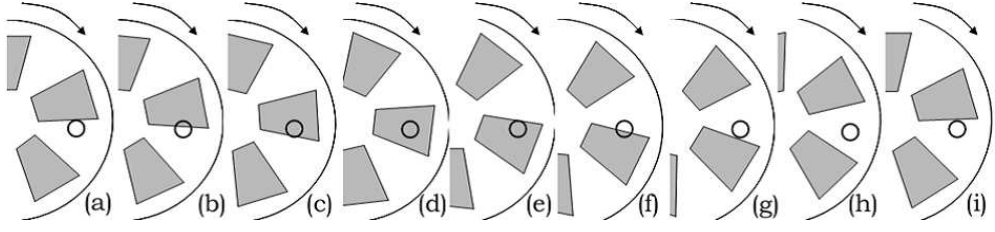}
\caption{The positions of rotating chopper holes and the window of the cell. This shows the time evolution during one time period $T$. The chopper has 5 holes and so we have 5 IR pulses during one rotation of chopper.}
\end{center}
\end{figure}

\section{Solution}

The initial condition of differential equation (\ref{eq:base}) is given by

\begin{equation}
\bar{\theta} (0)= \theta_{env}.
\end{equation}

Then the general solution is obtained.

\begin{equation}
\bar{\theta}(t) = \theta_{env} + \frac{1}{\mathscr{C}} \int_0^t L(t') e^{- (t-t')/\tau} dt', \label{eq:sol}
\end{equation}

where we used the definition

\begin{equation}
\tau \equiv \mathscr{C} \mathscr{R}.
\end{equation}

Since the form of $L(t)$ is not simple, we first decompose one to constant part and oscillation part.


\begin{figure}[htbp]
\begin{center}
    \includegraphics[width=0.8\hsize,keepaspectratio]{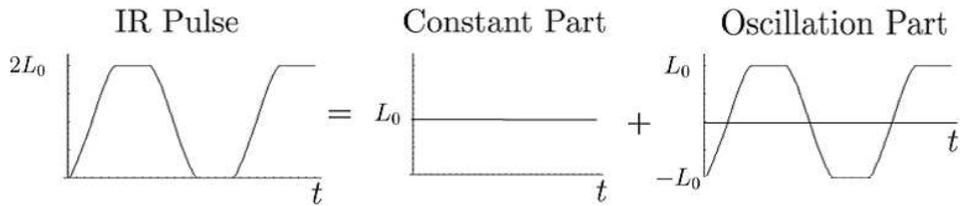}
    \caption{Decomposition of IR pulse}
\end{center}
\end{figure}
\begin{equation}
L(t) = L_0  (1 + f(t)),
\end{equation}
where $f(t)$ is the oscillating function with amplitude 1 and period $T$.
By putting this into (\ref{eq:sol}), we have
\begin{equation}
\bar{\theta}(t) = \theta_{env} +  \mathscr{R} L_0 (1-e^{-t/\tau}) + \frac{L_0}{\mathscr{C}}e^{-t/\tau} \int_0^t f(t') e^{t'/\tau} dt'.\label{eq:eq2}
\end{equation}

To perform the last integration
\begin{equation}
\int_0^t f(t') e^{t'/\tau} dt',
\end{equation}
we follow the steps below.

First,  $f(t)$ satisfies the relation.
\begin{equation}
f(t) = f(t+T), ~~~  \int_0^T f(t) dt =0. \label{eq:cond1}
\end{equation}

Second, we suppose the condition
\begin{equation}
T<< \tau=\mathscr{C} \mathscr{R},\label{eq:cond2}
\end{equation}
which we will prove later.

In general we can write $t=nT + \epsilon T$ ($n$:integer, $0 \le \epsilon \le 1$). Then we obtain

\begin{equation}
\int_0^t f(t') e^{t'/\tau} dt' = \int_0^{nT} f(t') e^{t'/\tau} dt' + \int_{nT}^{nT + \epsilon T} f(t') e^{t'/\tau} dt'.
\end{equation}
\\

By using $e^{Ty/\tau} \cong 1~(0\le y \le 1)$ from (\ref{eq:cond1}) and (\ref{eq:cond2}), we obtain

\begin{equation}
\int_0^{nT} f(t') e^{t'/\tau} dt'  = \frac{1-e^{nT/\tau}}{1-e^{T/\tau}}\int_0^T f(t') e^{t'/\tau} dt' \cong n \int_0^T f(t') dt' =0,
\end{equation}
and

\begin{equation}
\int_{nT}^{nT + \epsilon T} f(t') e^{t'/\tau} dt' \cong e^{nT/\tau}\int_0^{\epsilon T} f(t') dt' \cong e^{t/\tau} \int_0^t f(t') dt'.
\end{equation}

From these two relations we can change eq. (\ref{eq:eq2}) as follows.

\begin{equation}
\bar{\theta}(t) \cong \theta_{env} + \mathscr{R} L_0 (1-e^{-t/\tau}) + \frac{L_0}{\mathscr{C}}\int_0^t f(t') dt'.\label{eq:eq4}
\end{equation}

The physical meaning is the following.
The second term in r.h.s. shows the temperature change due to constant part.
At $t << \tau$, temperature grow up linearly with gradient  $L_0 \mathscr{R} /\tau$, but soon one reaches to the constant value at $t >> \tau$. This means the energy balance between IR incoming energy and outgoing energy by thermal diffusion.
\begin{equation}
\frac{\bar{\theta}-\theta_{env}}{\mathscr{R}}=L_0.
\end{equation}

On the oscillation part (third part of r.h.s.), because of the rapid oscillation satisfying $T << \tau$, 
the heat diffusion to the outside of cavity cannot chase the rapid change.
Then the integration of luminosity determines the temperature amplitude.
This is quite important condition to measure the IR absorption rate of gas correctly.

\section{Compared to experiment}

For our approximation to be valid, we need the condition $T << \tau$.
First we estimate the relaxation time $\tau$ for our apparatus.
The heat capacity $\mathscr{C}$  is calculated to be $4.06 \times 10^{-2}$ J/K from the sample cell volume of $3.14 \times 10^{-5}$ m$^3$ and the density of $1.98$ kg/m$^3$ and isovolumetric specific heat of $28.8$ J/(mol $\cdot$ K) for CO$_2$.
The thermal resistance $\mathscr{R}$ (See appendix) is calculated to be $27.4 \sim 54.8$ K/W from the cylinder cell's length $1.00 \times 10^{-1}$ m, the gas's heat conductivity of $\kappa = 1.45 \times 10^{-2}$ W/(m $\cdot$ K), and $b/b_0 = 0 \sim 1$.Then we obtain $\tau = 1 \sim 2 ~ \mbox{s}$, for $b/b_0 = 1 \sim 0$.
We work with frequency $1/T = 100 \sim 10000\mbox{Hz}$, that is,  $T = 10^{-4} \sim 10^{-2}$.
Then we always have $T << \tau$ as good approximation.

\subsection{Form of sound wave}
The IR pulse looks like trapezoidal wave but its integration is very close to a sinusoidal function.
This is shown in figure 6 theoretically and in figure 7 experimentally.
Because the rising and falling part of trapezoidal wave is expressed by linear function, and so its integrated form is parabolic, and the flat part of trapezoidal wave becomes linear function after the integration. The combination of these two functions look like the sinusoidal function.
Of course this is approximately true, and the obtained temperature oscillation includes higher harmonics.
On this point we will comeback later.


\begin{figure}
\centerline{\includegraphics[width=10cm]{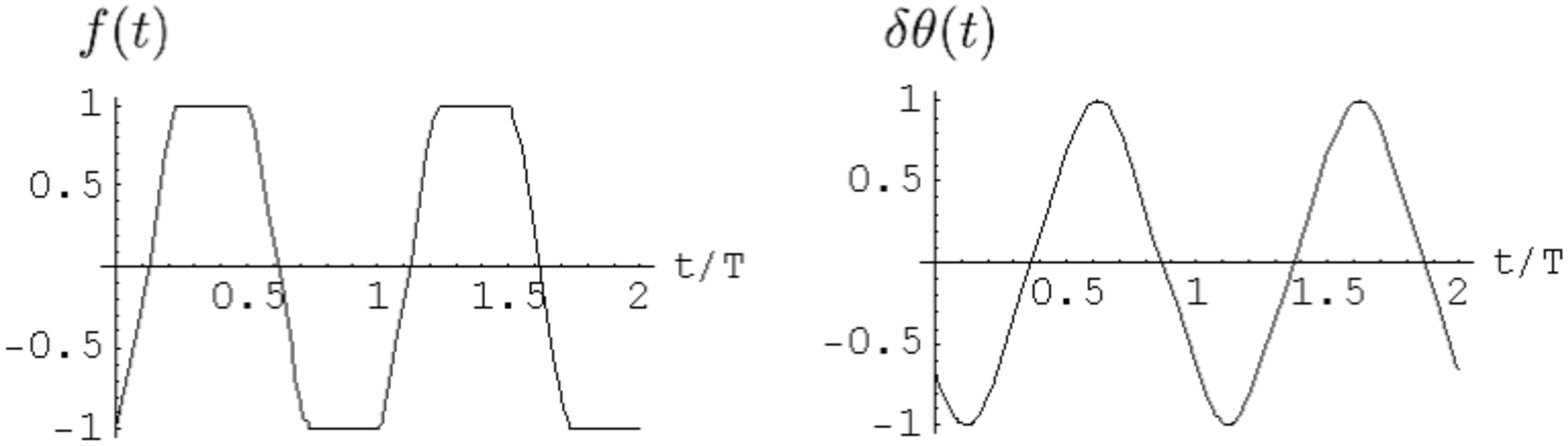}}
\caption{$f(t)$ and its integrated form as temperature oscillation, where amplitude is normalized. $\mu/T =0.24$ }
\end{figure}


\begin{figure}
\centerline{\includegraphics[width=4cm]{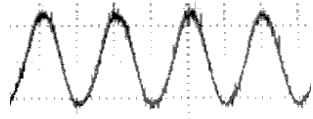}}
\caption{PA signal of CO$_2$ with vertical scale of 100 mV/div and horizontal scale of 10 ms/div.}
\end{figure}

\subsection{Frequency dependence of sound amplitude}

The normalized oscillating luminosity $f(t)$ can be expanded like

\begin{equation}
f(t) = \sum_{n=1}^\infty a_n \sin\{n \omega (t- \mu/2)\},
\end{equation}
where $\omega$ is the IR modulation frequency.
\begin{equation}
\omega = \frac{2\pi}{T}.
\end{equation}

Furthermore from the symmetry
\begin{equation}
f(t+\mu/2+T/4) = f(-t+\mu/2+T/4),
\end{equation}
we have
\begin{equation}
a_{2n}=0.
\end{equation}

By using equation (\ref{eq:eq4}), we obtain oscillation part of temperature.

\begin{equation}
\delta \theta (t) = - \frac{L_0}{\omega \mathscr{C}}\sum_{n=0}^{\infty} \frac{a_{2n+1}}{(2n+1)} 
\cos\{ (2n+1) \omega (t-\mu/2)\}.\label{eq:eq5}
\end{equation}

By using the state equation for an ideal gas under the condition of constant volume, we obtain pressure oscillation.

\begin{eqnarray}
\delta P(t) &=&  -\frac{A}{\omega} \sum_{n=0}^{\infty} \frac{a_{2n+1}}{(2n+1)}
\cos\{\omega (2n+1) (t-\mu /2)\},~ \label{eq:sound1}
\end{eqnarray}

with
$$A=\frac{z R_g}{V} \frac{L_0}{\mathscr{C}},$$
where $z$ is the number of moles of the gas, $R_g$ is the gas constant, and $V$ is the inner volume of the cell. The amplitude $| \delta P |$ is given as

\begin{equation}
|\delta P| =  - \delta P (t=\mu/2) = \frac{A}{\omega} \sum_{n=0}^{\infty} \frac{a_{2n+1}}{(2n+1)}. \label{eq:amplitude1}
\end{equation}

The sound amplitude is obviously proportional to the inverse of modulation frequency $\omega$.


\begin{figure}
\centerline{\includegraphics[width=7cm]{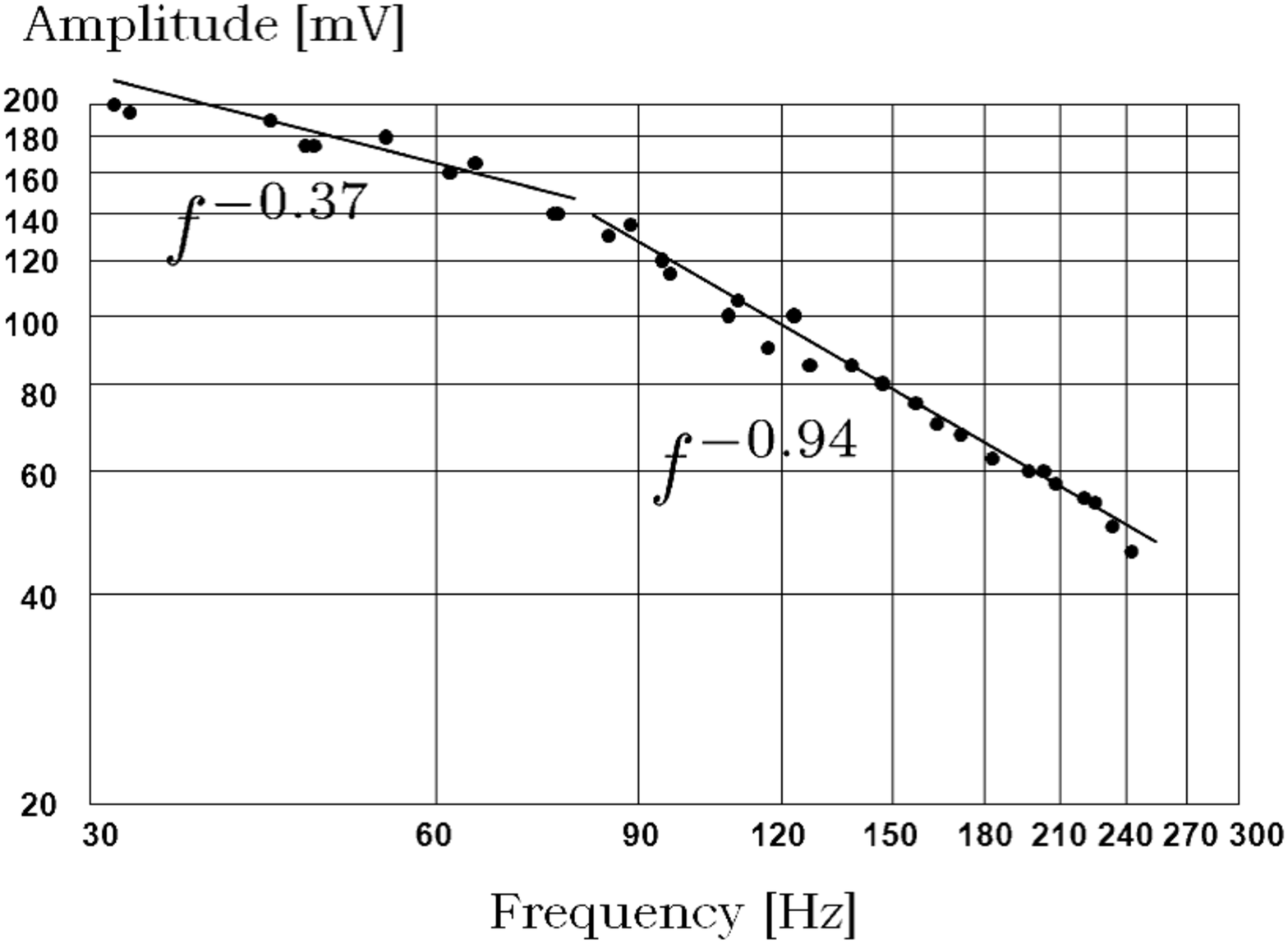}}
\caption{Relation between PA signal amplitude and IR modulation frequency.}
\end{figure}

In the experiment, the log-log plot of amplitude versus frequency is given in Figure 8.
Although the frequency dependence is expressed as $f^{-0.94}$ above 100 Hz, which is  consistent with the theory with 6\% error,  it is expressed as $f^{-0.37}$ below this frequency. 
The inconsistency in the low-frequency region can be ascribed to the frequency characteristics of the microphone and amplifier in our system.  
The sensitivity of the microphone and amplifier increases with frequency up to about 100 Hz and it is almost constant in the region of 100-10000Hz. 
So we can conclude that the amplitude of PA signal is inversely proportional to the modulation frequency. 

\subsection{Form of $a_n$}

The function $f(t)$ is too much complicated and so we can not show the analytic form of $a_n$, however, $f(t)$ can be approximated by trapezoidal function. 
In this case we have simple form for $a_n$.

\begin{equation}
a_{2n+1} = \frac{8}{\pi (2n+1)^2 \nu} \sin\{\frac{(2n+1)\nu}{2}\}, ~~~ a_{2n}=0. ~~n = 0, 1, 2 \cdots, \label{eq:spector}
\end{equation}

where $\nu$ is the constant rate factor of IR pulse. 
\begin{equation}
\nu \equiv \frac{2\pi}{T}\mu = \mu \omega.
\end{equation}

Then we have rectangular pulse wave for $\nu=0$, and triangular pulse wave for $\nu=\pi$.
In practice, the value of $\nu$ is determined by the relative size of the light entrance window of the cell and the chopper.  As shown in Figure 3 (a), $\nu$ is expressed as 

$$\nu/(2\pi) = \mu/T = \phi_2/\phi_1.$$

 To be exact, the shape of the entrance window gives an secondary effect on the form of the IR input power $L(t)$.  When the entrance window has a shape of a sector, $L(t)$ has an exact trapezoidal form.  
However, the shape of $L(t)$ becomes a gsmooth trapezoidal wave" with round corners for a system having a circular entrance window as our case, which is shown in Figure 3. 
In this case, we can not obtain analytical solutions for the coefficients $a_n$, so that we need to evaluate them numerically.  We will discuss the characteristics of sound signals as well as the influence of factor $\nu$ in both cases in section 5.
\\

The $\nu$ dependence of the amplitude can be obtained as follows.
Since $L_0$ is proportional to the area of cell window $\pi b^2$ as seen in appendix,
 and $b^2 \propto \phi_2^2 \propto \nu^2$ for fixed $\phi_1$ from figure 3(a), 
$A$ has the $\nu$ dependence as $A(\nu) \propto \nu^2$.
Further we have the relation
$$\sum_{n=0}^{\infty} \frac{\sin\{\frac{(2n+1)\nu}{2}\}}{ (2n+1)^3}  \propto \nu (1-\frac{\nu}{2\pi}).$$
Then by using  (\ref{eq:amplitude1}) and (\ref{eq:spector}), 
we obtain the $\nu$ dependence of the amplitude.

\begin{equation}
|\delta P| \simeq  \nu^2 (1-~ \nu /(2\pi) )/ \omega,\label{eq:amplitude}
\end{equation}

\section{Higher Harmonics of PA Sound}

The observed PA signal is not an exact sinusoidal wave because it includes higher harmonics. 
In this section we evaluate the contribution of the higher harmonics.  
Let $p_n$ as the n-th harmonics amplitude in PA signal. 
According to Eq.  (\ref{eq:sound1}), we have for trapezoidal approximation,

\begin{eqnarray}
\mid p_1\mid &=& \frac{8 A}{\pi \nu \omega} \sin (\nu /2), \\
\mid p_3\mid &=& \frac{8 A}{\pi \nu \omega} \frac{\sin (3 \nu /2)}{3^3}.
\end{eqnarray}

From this relation we have
\begin{equation}
g(\nu) \equiv \mid p_3/p_1\mid = \frac{\mid 1+ 2\cos \nu \mid}{3^3}.
\end{equation}

Let us consider the form of function $g(\nu)$.
$g(\nu)$ is the decreasing function for $0<\nu<2\pi/3$  and increasing function  for $2\pi/3<\nu<\pi$.
The physical reason is the following.
The slope of rising edge of trapezoid changes from infinity to smaller value for larger $\nu$,
 and so the higher harmonics are suppressed ($g(\nu)$ decreases) in a region of small $\nu$. 
But in a region of large $\nu$, the peak edge becomes more sharply bent for larger $\nu$ 
and the higher harmonics enhance ($g(\nu)$ increases).
This explains the form of dashed line (theoretical value for a sector window) in Figure 9.

The solid line shows the theoretical values for a circular window due to the numerical calculation.
The qualitative property is the same as the case of sector window, 
but the position of singularity ($p_3=0$) changes to higher $\nu$.

For the experimental values of spectrum of PA sound with circular window,
the first and third harmonics appear clearly, but higher ones  are buried in the background noise. 
So we can compare only the value of $\mid p_3/p_1\mid$ between theory and experiment.
The change of $\nu$ can be controlled  by changing the radius of  entrance window of cell $b$ with fixed $b_0$ experimentally.
As shown in Figure 9, we have the well coincidence with the experimental values and theoretical curve. \cite{spectrum}
This shows the validity of our model.
For the experiment it is necessary to give a comment. 
Since the IR is not prepared as parallel rays, even the shadow part of chopper gains the IR.
To avoid such a situation, it is important to put chopper very close to the window of the cell, 
and putting the source of IR as far as possible. 


\begin{figure}
\centerline{\includegraphics[width=6cm]{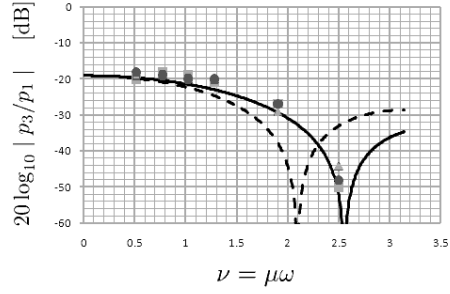}}
\caption{$\nu$-dependence of harmonics ratio $\mid p_3/p_1\mid$. 
The solid and dashed lines show the theoretical values for a circular and a sector window, respectively. 
The former corresponds to the smoothed trapezoidal wave, and the latter to the trapezoidal one.
The box, circle, and triangle shaped points show the experimental results with circular hole.}
\end{figure}

\section{Analogy with Electric Circuit}

To understand a new phenomenon the method of analogy is very convenient.
As we have already shown, our PAS is very close to the parallel CR circuit.


\begin{figure}
\centerline{\includegraphics[width=5cm]{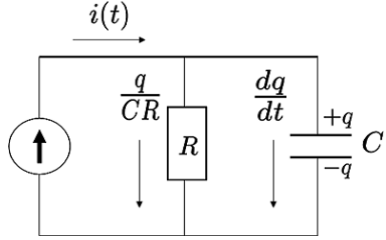}}
\caption{Parallel CR circuit}
\end{figure}

From Figure 10, we find the relation.

\begin{equation}
i = \frac{dq}{dt} + \frac{q}{CR} \label{eq:circuit}
\end{equation}

Let us show the correspondence relation between the variables in the electric circuit and those in the PAS.
\begin{itemize}
\item total electric current ($i$) $\to$ IR power ($L$),
\item electric resistance ($R$) $\to $ heat resistance ($\mathscr{R}$),
\item electric capacity ($C$) $\to$ heat capacity ($\mathscr{C}$),
\item electric charge of condenser ($q$) \\
$\to$ internal heat energy of gas ($U = \mathscr{C} (\bar{\theta}-\theta_{env})$),
\end{itemize}

From the above relations, we see that the resistive element corresponds to 
the heat resistance against the heat flow into the wall of gas cell,
 and the condenser corresponds to the heat capacity of the gas in the cell.

By using the analogy, we obtain the following relation from Eq. (\ref{eq:circuit}),
which is equivalent to Eq. (\ref{eq:base}).  Just like the electric circuit where the condenser is charged during a half cycle ($i(t)>0$) and the condenser is discharged into the resistive element in another half cycle ($i(t) = 0$), the temperature of the gas increases during the gas is irradiated by IR light and the temperature decreases by the energy flow during the rest of the irradiation, which explains well the IR input and the sound signal shown in Figure 4.

\section{Conclusion}
 We have introduced a simple model to deal with the photo acoustic phenomenon of a gas enclosed in a cell, whose energy gains and losses can be regarded to have the same relation as those in the parallel CR circuit.  
By using this model we can quantitatively explain the following three questions, 

\begin{itemize}
\item Why do we observe the sound signal having a sinusoidal like form when trapezoidal IR pulses are inputted into the sample gas ?
\end{itemize}

\begin{itemize}
\item How does the amplitude of the sound signal depend on the modulation frequency of the IR input?
\end{itemize}

\begin{itemize}
\item Why can we neglect the diffusion effect to measure the IR absorption rate in PAS ?
\end{itemize}

The analogy to the electric circuit leads us to understanding this phenomenon intuitively.  Though this model is simple, it is a very convenient way to explain our PAS for students.

\begin{acknowledgments}
The authors would like to thank Prof. Kuniaki Toyota in Hokkaido Institute of Technology for valuable discussion.
\end{acknowledgments}

\section{Appendix}

\subsection{INCIDENTAL IR BEAM}

Incidental IR beam power with unit [W/m$^2$] is expressed like

$$I(x,t) = I_0 e^{-\beta x} (1+f(t)),$$
where $f$ is the oscillating function with amplitude 1, 
and $\beta$ is the absorption rate per unit length.
Then the power density with unit [W/m$^3$]given to gas is

$$-\frac{d I}{dx} = \beta I_0 e^{-\beta x} (1+f(t)).$$

Some of their energy is re-emitted and do not contribute to increase the local temperature.
By using probability of radiation-less transition $\alpha$, we have an effective incident power density,

$$S = - \alpha \frac{d I}{dx} = \alpha \beta I_0 e^{-\beta x} (1 + f(t)).$$

We then define the net energy absorption per second [W],
$$L(t) \equiv \int S ~d^3x =  L_0 (1+f(t)),$$
with
$$L_0 \equiv \alpha  \pi b^2  I_0 (1-e^{-\beta l_0}),$$
where $l_0$ is the length of gas cell, and $\pi b^2$ is the area of cell window.

\subsection{BASIC EQUATION, AND ITS REDUCTION TO MEAN FIELD EQUATION}

Our starting equations are,

\begin{equation}
\kappa \nabla^2  \theta - \rho_0 ~ c_p ~\frac{\partial\theta}{\partial t}  + S = - \frac{\partial P}{\partial t}, \label{eq:temp}
\end{equation}
and
\begin{equation}
\nabla^2 P - \frac{\rho_0}{P_0} \frac{\partial^2 P}{\partial t^2} = -\frac{\rho_0}{\theta_0} ~ \frac{\partial^2 \theta}{\partial t^2}.\label{eq:press}
\end{equation}

Each field  is composed of  ambient constant and fluctuation field as,
$$\theta = \theta_0 + \delta \theta (x,t), ~~~ P =  P_0 +  \delta P (x,t).$$
Then we define the mean field by

\begin{eqnarray}
\bar{\theta}(t) &\equiv&  \theta_0 + \frac{1}{V} \int_{\Sigma} \delta \theta (x,t) d^3x, \\
\bar{P}(t) &\equiv&  P_0 + \frac{1}{V} \int_{\Sigma} \delta P (x,t) d^3x,
\end{eqnarray}
where $\Sigma$ is the inner space of gas cell, and $\partial \Sigma$ means its boundary hereafter.
To obtain the equation for these mean fields, we integrate Eqs. (\ref{eq:temp}) and (\ref{eq:press}) .

\begin{equation}
\kappa \int_{\partial \Sigma} \vec{\nabla}  \theta \cdot d\vec{\sigma} - \rho_0 V ~ c_p ~\frac{\partial \bar{\theta}}{\partial t}  + L = - V \frac{\partial \bar{P}}{\partial t}, \label{eq:mean}
\end{equation}
and
\begin{equation}
\int_{\partial \Sigma} \vec{\nabla} P \cdot d\vec{\sigma} - \frac{\rho_0 V}{P_0} \frac{\partial^2 \bar{P}}{\partial t^2} = -\frac{\rho_0 V}{\theta_0} ~ \frac{\partial^2 \bar{\theta}}{\partial t^2}.\label{eq:pressure}
\end{equation}

On the wall of cell,  the normal speed of gas fluid $v_n$ equals zero.

\begin{equation}
0= v_n = - \frac{1}{\rho_0} \int \vec{\nabla} P \cdot \vec{n} ~dt = \frac{i}{\rho_0 \omega} ~ \vec{\nabla} P \cdot \vec{n} ,
\end{equation}
where the linearized Euler equation $\dot{v} = -\nabla P/\rho$ is used, $\vec{n}$ is the normal unit vector to surface of cell, 
and oscillation with modulation frequency $\omega$ is supposed.
Therefore the first term of left hand side of Eq. (\ref{eq:pressure}) equals zero, 
and so we have after once integration.

\begin{equation}
\frac{\partial \bar{P}}{\partial t} = \frac{P_0}{\theta_0} ~ \frac{\partial \bar{\theta}}{\partial t}.
\end{equation}

Then together with Eq. (\ref{eq:mean}), we obtain
\begin{equation}
\kappa \int_{\partial \Sigma} \vec{\nabla}  \theta \cdot d\vec{\sigma} + L 
= \mathscr{C} ~\frac{\partial \bar{\theta}}{\partial t}, \label{eq:1}
\end{equation}
where the isovolumetric heat capacity $\mathscr{C} = \rho_0 V c_p -Nk$, with $N$ the number of gas particles, 
$k$ the Boltzmann constant.

Now the outgoing heat flow $\vec{J}$ is given by

$$ \vec{J} = - \kappa \vec{\nabla} \theta.$$

on the surface of cell inside.
We hope to express the heat going out of cell $\int \vec{J }\cdot d\vec{\sigma}$
by using $\bar{\theta}$ but not local function $\theta(\vec{x},t)$.

The usual theories for PAS \cite{theory} approximates the coordinate dependence 
only to the traveling direction of light $x$, however, the heat flow goes mainly to the radial direction
 in our PA sample cell consisting of a metal oblong tube and two KBr windows as shown in Figure 1.
So we approximate our heat flow  $\vec{J}$ going to the radial direction $r$ but not to $x$ direction instead.
Next we calculate the temperature distribution for static case and
 consider the relation between $\bar{\theta}$ and temperature gradient at the inner surface of cell.

Let us consider the static solution of Eq. (\ref{eq:temp}) with constant $S=L_0/(\pi b^2 l_0)$ for $r\le b$ and $S=0$ for $b < r \le b_0$.
$$ \kappa \nabla^2 \theta + S =0.$$

The boundary condition is as follows.
$\partial_r \theta =0$ at $r=0$, and $\theta = \theta_{env}$ at $r=b_0$, where $b_0$ is the inner radius of cell cylinder.
Since the thermal diffusibility of the metal wall of our sample cell is high, 
we treat the temperature at the inside surface of the wall as a constant value of $\theta_{env}$.
Then the solution of static equation is as follows.
First the inner solution
$$\theta_{in}(r) = \frac{L_0}{4\pi b^2 l_0 \kappa} (b^2 - r^2) + \theta(b),$$
for $r\le b$. Second the outer solution
$$\theta_{out}(r) = \theta(b) - \frac{\theta(b) - \theta_{env}}{\ln(b_0/b)} \ln (r/b),$$
for $b < r \le b_0$.
To obtain $\theta(b)$, additional boundary condition is necessary.
$$\kappa \frac{\partial \theta_{in}}{\partial r}\mid_{r=b} = \kappa \frac{\partial \theta_{out}}{\partial r}\mid_{r=b}.$$
This gives solution.
$$\theta(b)=\theta_{env} + \frac{L_0}{2\pi l_0 \kappa}\ln(b_0/b).$$

and $\bar{\theta}$ is
\begin{eqnarray}
\bar{\theta} &=& \frac{1}{\pi b_0^2} \int \theta(r) r d\phi dr = \frac{2}{{b_0}^2} (\int_0^b r \theta_{in} dr + \int_b^{b_0} r \theta_{out} dr) \nonumber\\
&=& \theta_{env} + \frac{L_0 \{2-(b/b_0)^2\}}{8 \pi l_0 \kappa}. \label{eq:relation}
\end{eqnarray}
Note that from $L_0 \propto b^2$, $\bar{\theta} \to \theta_{env}$ as $b \to 0$.

The outgoing heat energy from gas cell per second is the same as net energy absorption per second from IR  in static case. So we have $\int \vec{J }\cdot d\vec{\sigma} = L_0$.
Therefore we interpret Eq. (\ref{eq:relation}) as
\begin{equation}
\bar{\theta}-\theta_{env} = \mathscr{R}  \int \vec{J }\cdot d\vec{\sigma}  ,\label{eq:2}
\end{equation}
with definition of heat resistance
\begin{equation}
\mathscr{R} \equiv \frac{2-(b/b_0)^2}{8 \pi l_0 \kappa}. \label{eq:3}
\end{equation}

We suppose Eq. (\ref{eq:2}) holds even for the case of non static temperature.
Then from Eqs. (\ref{eq:1}) and (\ref{eq:2}), we obtain Eq. (\ref{eq:base}).


\begin{thebibliography}{99}
\bibitem{Kaneko} F. Kaneko, H. Monjushiro, M. Nishiyama, and T. Kasai, Journal of Chemical Education, 3b2, Ver. 9, (2009) ed-2008000446

\bibitem{history} A .G. Bell, Am. J. Sci. {\bf 20}, 305 (1880); J. Tyndall, Proc. R. Soc. London {\bf 31}, 307 (1881); 
W. C. Roentgen, Philos. Mag. {\bf 11}, 308 (1881); A. G. Bell, ibid {\bf 11}, 510 (1881)

\bibitem{GH} Intergovernmental Panel on Climate Change. Climate Change 2007: The Physical Science Basis; Cambridge University Press: Cambridge (2007)

\bibitem{theory} F. A. McDonald, and G. C. Wetsel, Jr. , J. Appl. Phys. {\bf 49}, (4) p 2313- p 2322 (1978); F. A. McDonald, Am. J. Phys. {\bf 48} (1) (1980);  Rosencwaig, Allan. Photoacoustics and Photoacoustic Spectroscopy. New York: John Wiley and Sons (1980); Philip M. Morse and K. U. Ingard, ``Theoretical Acoustics" Princeton Univ. Press, Princeton New Jersey, (1968).

\bibitem{spectrum} We have analyzed the form of PA signal with its spectrum amplitude by using Tektronix TDS1012B.
\end{thebibliography}
\end{document}